# Superconductivity in amorphous $Re_xZr$ (x ≈ 6) thin films

Surajit Dutta[a], Vivas Bagwe[a], Gorakhnath Chaurasiya[b], A. Thamizhavel[a], Rudheer Bapat[a], Pratap Raychaudhuri[a], Sangita Bose[b*]

[a] *Department of Condensed Matter Physics and Material Science, Tata Institute of Fundamental Research, Homi Bhabha Road, Colaba, Mumbai 400005, India.*

[b] *School of Physical Sciences, UM-DAE Center for Excellence in Basic Sciences, University of Mumbai, Kalina Campus, Santacruz E, Mumbai 400098, India*

*We report the growth, characterization and superconducting properties of a new amorphous superconductor, $Re_xZr$ ( $x ≈ 6$ ), in thin film form. Films were grown by pulsed laser deposition with the substrate kept at room temperature. Films with thickness larger than 40 nm showed a superconducting transition temperature ($T_c$) of 5.9 K. Superconducting properties were measured for films with varying thickness from 120 to 3 nm. The normal state resistance scales linearly with inverse of thickness. The transition temperature, critical field, coherence length, penetration depth and superconducting energy gap changes marginally with decreasing film thickness down to 8 nm. Scanning tunneling spectroscopy and penetration depth measurements provide evidence for a single gap strong coupling s-wave superconductor. Magneto-transport properties indicate a rich magnetic field-temperature phase diagram with the possibility of vortex liquid phases existing over a large fraction of the mixed state.*

---

[*] sangita@cbs.ac.in

I. Introduction

Crystalline $Re_6Zr$ is a non-centrosymmetric superconductor (NCS) in the bulk form with a superconducting transition temperature *($T_c$)* of 6.7 K[1,2]. It has a α-Mn cubic structure which favors a time reversal symmetry (TRS) breaking with singlet-triplet mixing[1,3]. While bulk measurements on both single crystal and polycrystals showed the existence of a fully gapped single band[4,5,6], muon spin rotation measurements indicated signatures of time reversal symmetry breaking. Point contact Andreev reflection measurements also showed the presence of multiple gaps suggesting unconventional modification of superconductivity close to the surface[7]. These observations motivated us to attempt the synthesis of this superconductor in thin films form where the surface to bulk ratio can be controlled.

Metals and metallic alloys when deposited in the form of thin films sometimes stabilize in amorphous phase. Our attempts of synthesizing $Re_6Zr$ thin films resulted in the synthesis of a new amorphous superconductor (AS) with composition close to its crystalline counterpart. The discovery of superconductivity in the amorphous phase[8] dates back to the early 1950's when quench condensed films of amorphous Bi was reported to be superconducting with a $T_c$ of 6.1 K[9]. In contrast, crystalline Bi has only recently been shown to be a superconductor[10] with $T_c$ of 0.53 mK. Thereafter, several amorphous superconductors have been discovered some of them showing $T_c$ higher than their crystalline counterparts. Typically, AS-s are strongly coupled Type II superconductors, with the superconducting energy gap ($\Delta(0)$) exceeding the Bardeen-Cooper-Schreiffer (BCS) ratio, ($\Delta / k_B T_c$) ~1.76. Due to their very short electronic mean free path these materials form good model systems to study the role of disorder on superconductivity[11,12,13,14,15,16,17,18,19,20,21]. The emergence of novel vortex phases in AS has also been another focal point of interest[22,23,24,25,26]. From the standpoint of applications AS has been used for fabricating superconducting detectors[27,28,29] and vortex memory devices[30].

In this article, we report the growth and superconducting properties of amorphous $Re_xZr$ ($x \approx 6$) thin films grown by pulsed laser deposition (PLD). We synthesized amorphous $Re_xZr$ (which we will henceforth refer to as $a$-ReZr) film up to a thickness ($t$) of 120 nm. Interestingly, all films with 3 nm $\leq t \leq$ 120 nm were superconducting with the transition temperature approaching a limiting value, $T_c \sim 5.9$ K, for $t \geq 40$ nm. The superconducting energy gap, $\Delta(T)$ follows the BCS temperature variation. However, $\Delta(0)$ obtained from both penetration depth and scanning tunneling spectroscopy (STS) measurements show that the superconductor in the strong coupling limit. Furthermore, the superfluid stiffness, $J$, estimated from penetration depth show that phase fluctuations play a negligible role in the superconducting properties down to $t = 5$ nm.

## II. Experimental Details

$a$-ReZr films of different thickness ($t$ = 3 - 120 nm) were grown on surface oxidized Si (100) (oxide layer thickness $\sim$ 200 nm) substrates by pulsed laser deposition. The target with composition $Re_6Zr$ was made using arc-melting starting from stoichiometric ratios of high purity (99.99%) Rhenium powder and 99.999% Zirconium shots both from Alfa Aesar. Rhenium powder was first arc melted to form small Rhenium shots in a mono arc furnace. Zirconium shots were then taken along with the Rhenium shots in stoichiometric amounts (Re:Zr $\Rightarrow$ 6:1) and melted to form a button in a tetra arc furnace under a partial pressure of ultra-high pure Argon gas on a water cooled hearth. This button ( $\sim$1.1 cm diameter ) was flipped several times and re-melted to obtain a homogenous mixture. It was characterized through x-ray diffraction (XRD) $\theta$ - $2\theta$ scans using monochromatic Cu K$\alpha$ source. XRD of the target showed the formation of the Re$_6$Zr crystalline α-Mn phase (Fig. 1(a)). Subsequently, the top surface was polished with fine emery paper to obtain a flat, smooth ablating surface. The films were grown by ablating the Re$_6$Zr target using a 248 nm excimer laser maintaining the substrate at room temperature. The deposition was carried out in vacuum of $1 \times 10^{-7}$ Torr. The laser spot was tightly focused on the target to obtain a high energy density, ~200 mJ mm$^{-2}$ per pulse, with repetition rate of 10 Hz. This high energy density was essential to maintain the stoichiometry of the film close to that of the target. The growth rate was ~1 nm/100 pulses. Films with varying thickness were grown by changing the number of laser shots. For $t > 10$ nm, the thickness was directly measured using a stylus surface profilometer. For thinner samples, it was estimated from the number of laser pulses using the calibration from thicker samples. The films were structurally characterized by XRD and the morphology was observed using a Nanosurf Atomic force microscope (AFM). Transmission electron microscopy (TEM) was performed using a Tecnai 20-200 microscope with LaB$_6$ filament operated at 200 kV. For TEM measurements, free-standing $a$-ReZr films were transferred on a Cu grid. This was done as follows. First $a$-ReZr films were deposited on surface oxidized Si substrates coated with a PMMA layer. The PMMA was dissolved by dipping the substrate in acetone and the free-floating film on acetone was then picked up on a Cu grid. It was verified that 20 nm thick film on PMMA coated substrate had the same $T_c$ (within 0.1 K) as the film of same thickness on surface oxidized Si substrate. To prevent surface oxidation, samples with $t < 10$ nm used for magneto-transport were covered with a 2-nm-thick protective Si layer. Magnetotransport measurements were performed in a $^4$He bath cryostat operating down to 1.8 K or a $^3$He cryostat operating down to 300 mK fitted with a superconducting solenoid with magnetic field up to 110 kOe. Recently, it has been shown that the transport properties of amorphous superconductors can be extremely susceptible to ambient electromagnetic radiation[31,32] (EMR). Therefore, to prevent any artifact from external EMR all electrical feedthroughs leading to the sample were fitted with low pass RC filters with cut-off frequency of 340 kHz. For magneto-transport as well as to measure the Hall voltage, the samples were deposited in the shape of Hall bar (Length: 1.3 mm and width = 0.35 mm). The London penetration depth ($\lambda$) was measured from the magnetic shielding response using a home-built setup up based on the low-frequency (31 kHz) two-coil mutual inductance technique[33,34]. In this technique, the film grown in a circular geometry of 8 mm diameter is placed in between a quadrupolar primary coil and a dipolar secondary coil such that the magnetic shielding response is given by the complex mutual inductance, $M = M' + iM''$ measured at 31 kHz. The penetration depth is obtained from the mutual inductance using the numerical procedure described in ref. 35. Details of this technique is given in earlier papers[36,37]. Penetration depth measurements were done either in a cryogen free system operating down to 2.8 K or in a $^4$He cryostat for measurements down to 2 K. STS measurements were performed using a homebuilt scanning tunneling microscope (STM) operating down to 450 mK and in magnetic fields up to 90 kOe[38]. Tunneling measurements were performed with Pt-Ir normal-metal tip. The tunneling conductance (G(V) = dI/dV) was measured using standard modulation technique. To determine the superconducting energy gap ($\Delta$), G(V) vs. V tunneling spectra were obtained at every point on a 32 × 32 grid over 200 nm ×200 nm area and the average spectra was analyzed to obtain $\Delta$. All STS measurement were performed on pristine surfaces by transferring the sample directly in an ultrahigh vacuum suitcase after deposition, and transferring in the STM without exposure to air.

**III. Results**

Fig. 1(a) shows the representative x-ray diffraction (XRD) θ – 2θ scans for *a*-ReZr films. We observe only broad humps around the Re$_6$Zr main peaks between 2θ of $40^0$ to $50^0$. High resolution TEM image captured on a 40 nm thick sample is uniform and featureless and selected area electron diffractions taken at several locations exhibit a diffuse ring pattern (Fig. 1(b)). All these observations confirm the amorphous nature of the film. AFM measurement shows smooth films with no granularity giving a surface roughness of ~ 1 - 1.5 nm (Figs. 1(c)-(d)). Energy dispersive x-ray (EDX) analysis of several films measured at various locations on the surface showed *x* in Re$_x$Zr to be in the range 5.9 - 6.1.

Fig. 2(a) shows the temperature variation of the sheet resistance ($R_s$) for films with different thickness (*t*). We observe that the sheet resistance at 9 K, $R_s^{9\,K} \propto \frac{1}{t}$, implying that the corresponding normal state resistivity, $\rho_N$, does not significantly change with disorder (inset of Fig. 2(a)). In addition, all films show a weak negative temperature coefficient of resistance coupled with large normal state resistivity ($\rho_N$) between 2 – 3 μΩ-m (at 9 K). This is indicative of a bad metal. Fig. 2(b) shows as expanded view of the superconducting transition. Fig. 2(c) shows the magnetic shielding response measured using the two-coil mutual inductance setup for films with different thickness. The superconducting transition temperature ($T_c$) was determined either from $R_s$-$T$ or from the magnetic shielding response. From $R_s$-$T$, $T_c$ was defined to be the temperature at which the resistance falls to 0.05% of the normal state value. This corresponds to the temperature at which persistent shielding currents become non-zero. Consequently, in the shielding response, we define $T_c$ as the temperature where *M'* drops to 99% of the normal state value. The two definitions give exactly the same value of $T_c$ when the measurement is performed on the same sample, as shown in Fig. 2(d). All films showed a reasonably narrow single peak in *M''* - *T*, close to $T_c$ showing that they are single phase. The variation of $T_c$ with thickness is shown in Fig. 2(e). The $T_c$ remained more of less constant to about 5.86 $\pm$ 0.06 K for thickness between 120 to 40 nm; the error in $T_c$ denotes statistical variation between different growth runs. Down to 8 nm, there was a gradual decrease in $T_c$ by about ~1 K. For films with thickness < 8 nm, $T_c$ decreases sharply, with the 3 nm film showing a $T_c$ of 2.3 K. In Fig. 2(f) we plot the magnetic field dependence of the Hall resistivity, $\rho_{xy} = \frac{V_H t}{I}$ (where $V_H$ is the Hall voltage, $I$ is the measurement current), for films with $t$ = 20 nm and $t$ = 40 nm. The carrier density, $n \sim 9.0(\pm 0.4) \times 10^{29}\ el\ m^{-3}$, was obtained from the Hall coefficient, $R_H = \frac{\rho_{xy}}{H}$, using the relation, $n = \frac{1}{eR_H}$. Using the free electron relations, we obtain the electronic mean free path ($l$),

$$l = \frac{m\, v_F}{n\rho_n e^2}, \quad v_F = \frac{\hbar k_F}{m} = \frac{\hbar}{m}(3\pi^2 n)^{1/3} \quad (1)$$

where $m$ is the electron mass, $e$ is the electron charge, and $v_F(k_F)$ is the Fermi velocity (wave-vector). We obtain the electronic mean free path, $l \sim 0.5$ Å (assuming the effective mass of the electron to be the free electron mass). This is smaller than that reported in $a$-MoSi[29], $a$-MoGe[39] and disordered NbN films[40]. The corresponding value of $k_F l \sim 1$ suggests that the films are strongly disordered.

Figs. 3(a) – (b) show the magnetic field variation of $R_s$ for the 40 and 5 nm thick films respectively measured at different temperatures. The magnetic field is applied perpendicular to the film plane. For the 5 nm thick sample, the magnetoresistance curves cross around 72 kOe, reflecting the weak negative temperature coefficient of resistance in the normal state. In the presence of magnetic field, the transitions get considerably broad. This field-induced broadening is often attributed to the emergence of a vortex liquid state at the field where resistance appears[41]. At a characteristic field, $H_I$ (called the Irreversibility field) the vortices become mobile, and the critical current vanishes. We define $H_I$ where $R_s$ is 0.05% of the normal state

value. The upper critical field, $H_{c2}$, is defined to be the field at which $R_s$ reaches 90% of the normal state value. Figs. 3 (c)-(e) we plot $H_I$ and $H_{c2}$ for 3 representative films. To determine the $H_{c2}(0)$ and hence the Ginzburg-Landau coherence length $\xi_{GL}(0)$ we use the following expression applicable in the dirty limit[42]

$$H_{c2}(0) = 0.693 * T_c \left(\frac{dH_{c2}}{dT}\right)_{T_c} \text{ and } \xi_{GL}(0) = \left(\frac{\Phi_0}{2\pi H_{c2}}\right)^{1/2} \quad (2)$$

$H_{c2}(0) \sim 11$ T for the 40 nm thick film and decreases to 6.5 T at 5 nm. The corresponding $\xi_{GL}(0)$ ranged between 5.5 – 7.1 nm (Fig. 3(f)).

Finally, we focus on two other crucial microscopic quantities, $\Delta$ and $\lambda$. In Figs. 4(a) and 4(c), we plot the normalized tunneling conductance, $G_N(V) = \frac{G(V)}{G(6\ mV)}$, as a function of bias voltage on the 20 nm and 5 nm thick film. $\Delta(T)$ is extracted at all temperatures by fitting these spectra to the tunneling equation[43],

$$G(V) = \frac{d}{dV}\left[\frac{1}{R_N}\int_{-\infty}^{+\infty} N_S(E)N_N(E-eV)(f(E)-f(E-eV))dE\right] \quad (3)$$

where $N_S(E)$ and $N_N(E)$ are the normalized density of states for the superconducting and normal metal respectively, $f(E)$ is the Fermi-Dirac distribution function and $R_N$ is the resistance of the tunnel junction for $V \gg \Delta/e$. $N_S(E)$ is given by the expression $N_S(E) = Re\left\{\frac{|E-i\Gamma|}{\sqrt{|E-i\Gamma|^2-\Delta^2}}\right\}$ where $\Delta$ is the superconducting energy gap and $\Gamma$ is an additional phenomenological parameter[44] which accounts for all non-thermal sources of broadening. Fig. 4(b) and (d) show the temperature variation of $\Delta$ and $\Gamma$ obtained from the fits. The temperature dependence of $\Delta$ follows the BCS variation within the experimental accuracy. However, $\frac{\Delta(0)}{k_B}$ is much larger than the BCS value suggesting that it is a strong coupling superconductor, as also observed for other reported AS[29,39,21]. Fig. 4(e) shows $\lambda^{-2}$ as a function of temperature for films with different thickness. We fit the data with the dirty limit BCS expression[43],

$$\frac{\lambda^{-2}(T)}{\lambda^{-2}(0)} = \frac{\Delta(T)}{\Delta(0)} tanh\left[\frac{\Delta(T)}{2k_BT}\right] \quad (4)$$

using Δ(0) and λ(0) as fit parameters and assuming a BCS temperature variation of Δ(*T*). The best fit values of Δ(0) and λ(0) are shown in Fig. 4(f). For 5 nm and 20 nm thick samples Δ(0) matches with the value obtained from tunneling measurements. It is interesting to note that within error bars *λ(0)* does not show any appreciable change with thickness. The relatively large error bar on the absolute value of $\lambda^{-2}$ is primarily due to the corresponding uncertainty in thickness measurement.

## IV. Discussions and Conclusion

Having ascertained that the thin films of *a*-ReZr behave like a BCS superconductor, we can try to understand the mechanism for the variation of $T_c$ with thickness in this system. Since our films are in the strong disorder limit, we need to consider two distinct, but not mutually exclusive mechanisms. In the first mechanism, with increase in disorder screening gets less effective, and the electron-electron Coulomb repulsion partially compensates the phonon mediated pairing interaction thereby causing $T_c$ to decrease. The theory of this effect was worked out for thin films by Finkelstein, which predicts a correlation between $T_c$ and the sheet resistance $R_s$ given by[17],

$$\frac{T_c}{T_{c0}} = \exp(\gamma)\left(\frac{1-X}{1+X}\right)^{1/\sqrt{2r}} \quad, \quad r = \frac{e^2}{2\pi^2\hbar}R_s \, , \quad X = \frac{\sqrt{r/2}}{\frac{r}{4}+\frac{1}{\gamma}}, \quad \gamma = ln(\hbar/\tau T_{c0}k_B) \quad (5)$$

Here $T_{c0}$ is the limiting value of $T_c$ for large thickness. In Fig. 5(a) we show that we can fit our data with eqn. (5) reasonably well, with $T_{c0} \sim 6.5$ K and $\gamma \approx 8$; the value of $\gamma$ is comparable to that obtained for a-MoSi[29] and a-MoGe[39] films. The second mechanism is the destruction of superconductivity from thermal phase fluctuation. The resilience of the superconductor against phase fluctuations is given by the superfluid stiffness, $J$, which can be estimated using the formulae (in SI units) [45,34],

$$J = \frac{\hbar^2 n_s a}{4m} , \quad n_s = \frac{m}{\mu_0 e^2 \lambda^2} \quad (6)$$

where $\mu_0$ is the permeability of vacuum and $a = \min\{t, \xi_{GL}(0)\}$. Phase fluctuations start playing an important role when $J(0) \lesssim \Delta(0)$. In our samples down to $t = 5$ nm $J(0)$ is larger than $\Delta(0)$ (Fig. 5(b)). Therefore, we do not expect phase fluctuations to play a significant role. We also do not observe any characteristic signature of phase fluctuations[46], such as the existence of a pseudogap state or a linear decrease of the superfluid density. However, it has been recently shown that in a-MoGe there is a crossover from the first mechanism to the second[21] when the thickness becomes ~ 2 nm. The possibility of such a cross-over in a-ReZr will have to be investigated in more detail in future.

Finally, we would like to note, recent experiments in a-MoGe show that a hexatic vortex fluid state[26,32] encompasses a large fraction of the mixed state and exists down to very low temperatures. The large difference between $H_I$ and $H_{c2}$ (Fig. 3(c)-(e)) in a-ReZr suggest the existence of vortex fluid phase over a large region in the H-T parameter space here as well. Therefore, it would be interesting to explore the nature of this vortex fluid state through real space STS imaging in future studies.

In summary, we were successful in growing thin films of a new amorphous superconductor, Re$_x$Zr ($x \approx 6$), by pulsed laser deposition. Measurement of the superconducting energy gap and penetration depth show that this is a conventional strong coupling Type II superconductor, with upper critical field exceeding 100 kOe. The ability to grow homogeneous high quality thin films at room temperature might make this material interesting for potential applications. Further investigation will be needed to determine the range of $x$ over which Re$_x$Zr remains amorphous and superconducting.

**Acknowledgements**

This work was financially supported by the Department of Atomic Energy, Government of India and Department of Science and Technology, Government of India (Grant Nos. SERB/EMR/2017/0007774 and12-R&D-TFR-5.10-0100). We thank John Jesudasan for technical help.

**Author contribution**

SB conceived the problem. SB and PR planned the methodology and supervised the experiments. VB and AT synthesized and characterized the samples. RB performed the TEM measurements. SD performed the transport and STM measurements and analyzed the data. GC performed AFM measurements. GC and SD performed penetration depth measurements and GC analyzed the data. The manuscript was written by SB and PR with inputs from all authors. All authors commented on the manuscript.

**Figure Captions**

**Figure1|** (a) X-ray diffraction $\theta$–$2\theta$ scans of the $Re_6Zr$ target (left axis) and *a*-ReZr films (right axis). (b) High resolution TEM image of 40 nm thick film and representative selective area electron diffraction pattern (inset). (c) and (d) AFM topographic image of two *a*-ReZr films; one particulate can be observed within the field of view in the 20 nm sample.

**Figure2|** (a) Temperature dependence of $R_s$ for *a*-ReZr films with different thicknesses. (*inset*) $R_s$ at 9 K and corresponding normal state resistivity, $\rho_N$, as a function of $1/t$; (b) expanded view of $R_s$ close to the superconducting transition. (c) The real and imaginary part of the magnetic shielding response, $M = M' + iM''$, as a function of temperature; (d) Comparison of the temperature dependence of $R_s$ and $M'$ on the same sample with $t = 20$ nm, close to the superconducting transition; the dotted line denotes the superconducting transition temperature, $T_c$. (e) Variation of $T_c$ with $t$ measured from $M$-$T$ and $R_s$-$T$. (f) $\rho_{xy}$ as a function of magnetic field measured at 25 K for films with 20 nm and 40 nm thickness.

**Figure 3|** Variation of $R_s$ with magnetic field for two films with thickness, (a) 40 nm and (b) 5 nm. Variation of the irreversibility field $H_I$ and the upper critical field $H_{c2}$ with $T$ for the films with thickness (c) 5 nm (d) 20 nm and (e) 40 nm. (f) Variation of $H_{c2}(0)$ and $\xi_{GL}(0)$ with film thickness.

**Figure 4|** (a) $G_N(V)$ vs. $V$ tunneling spectra at various temperatures for the 20 nm thick *a*-ReZr thin film; each spectrum is the average of 1024 points over 200 nm × 200 nm area. The solid lines show the fit using eqn. 3. (b) Temperature variation of $\Delta$ and $\Gamma$ for the same film; the solid blue line is the fit to the BCS temperature dependence. (c) (d) same as (a) and (b) but for a 5 nm thick *a*-ReZr film. (e) Temperature variation of $\lambda^{-2}$ for films of different thickness; the solid lines

are fits with dirty limit BCS expression (eqn. 4). Representative error bars are shown for two thickness. (f) Variation of $\lambda(0)$ and $\Delta(0)$ with film thickness; for $\Delta(0)$, the red square are from the fit of $\lambda^{-2}(T)$ and the orange star are from tunneling.

**Figure 5|** (a) Dependence of $T_c$ on the $R_s$ at 9K; the solid line is the fit with the expression in eqn. 5. (b) Comparison of $\Delta(0)$ (blue circle) and $J_s(0)$ (green square) for films of different thickness; the lines are guide to the eye.

**Figure 1**

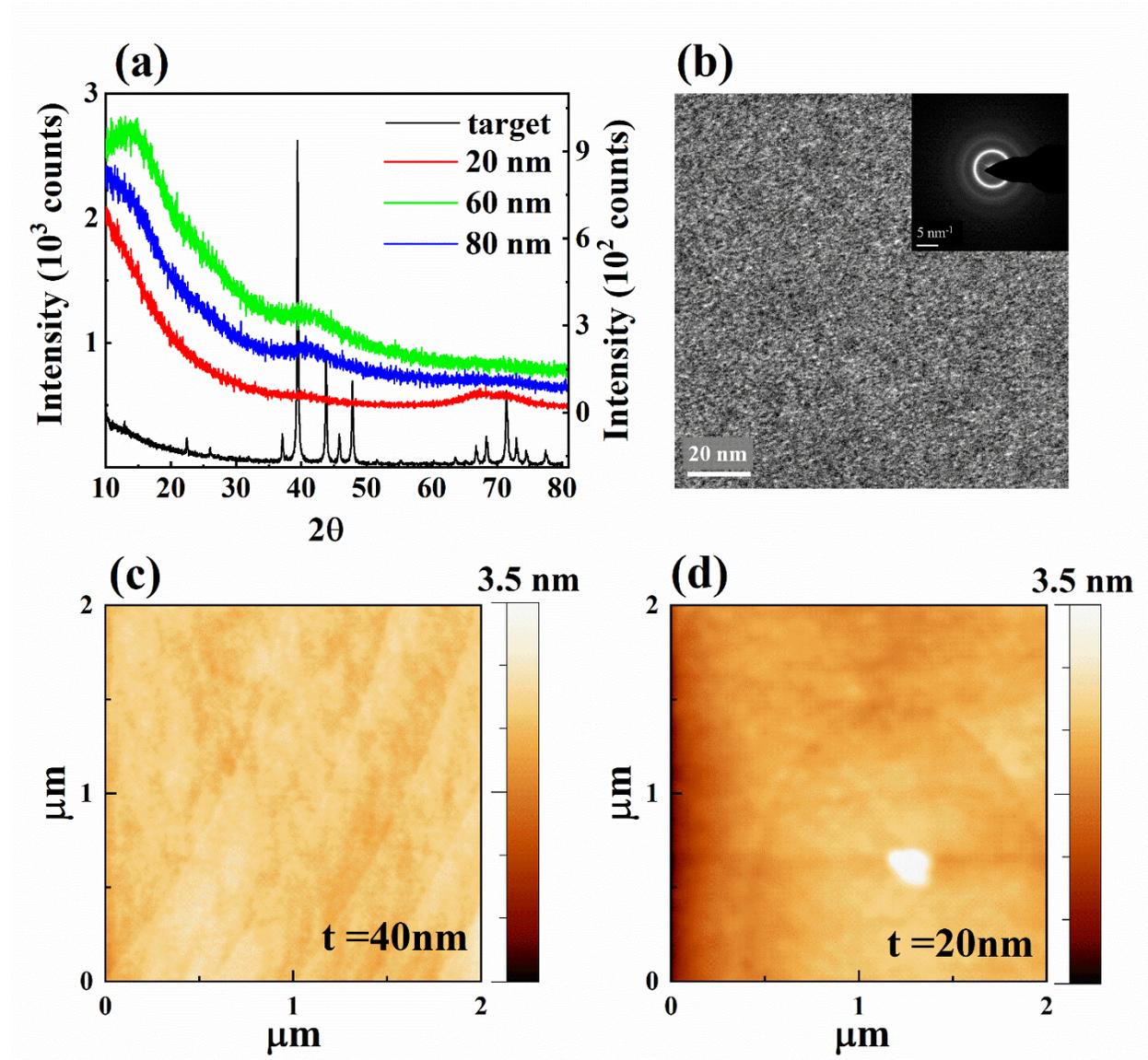

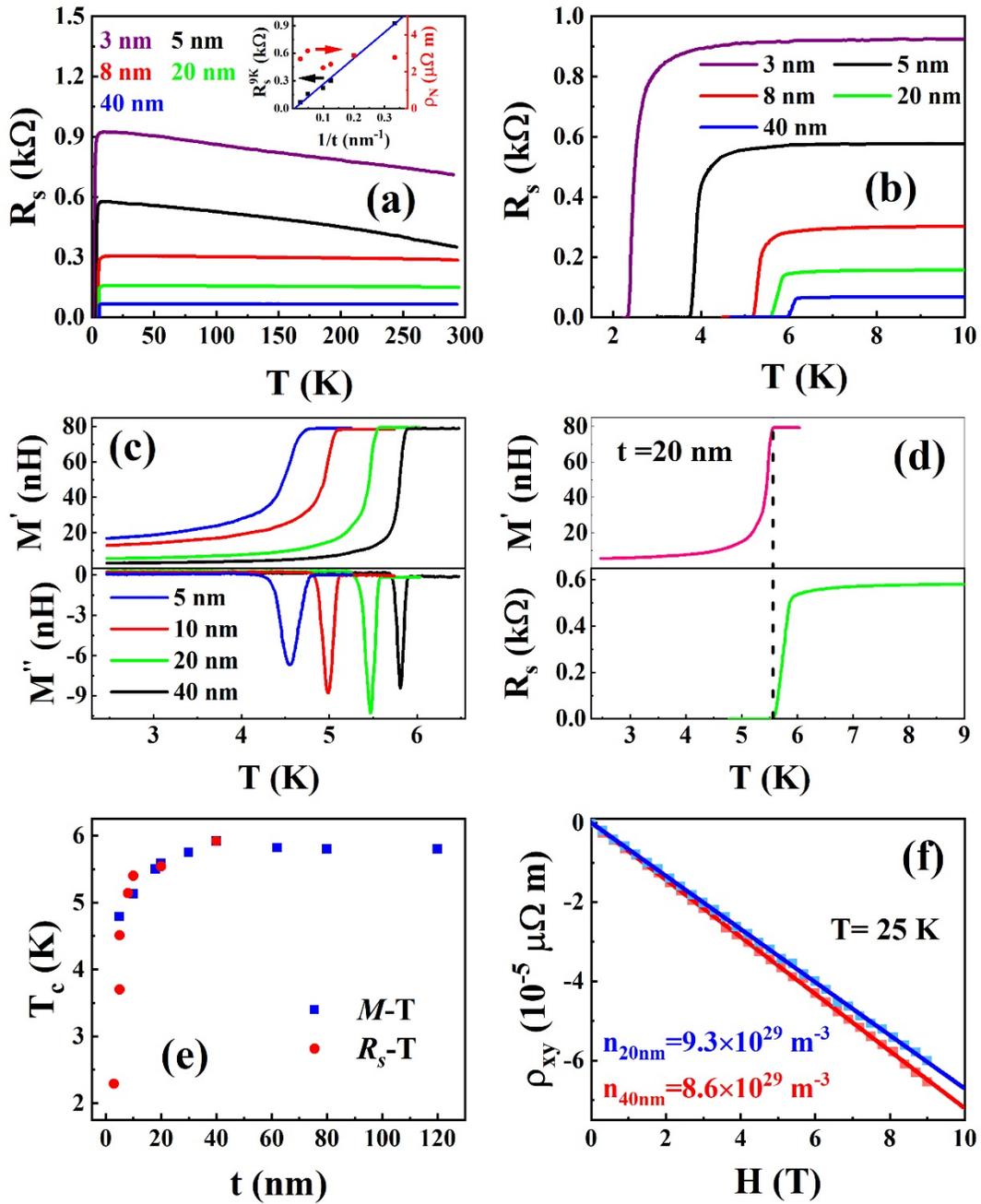

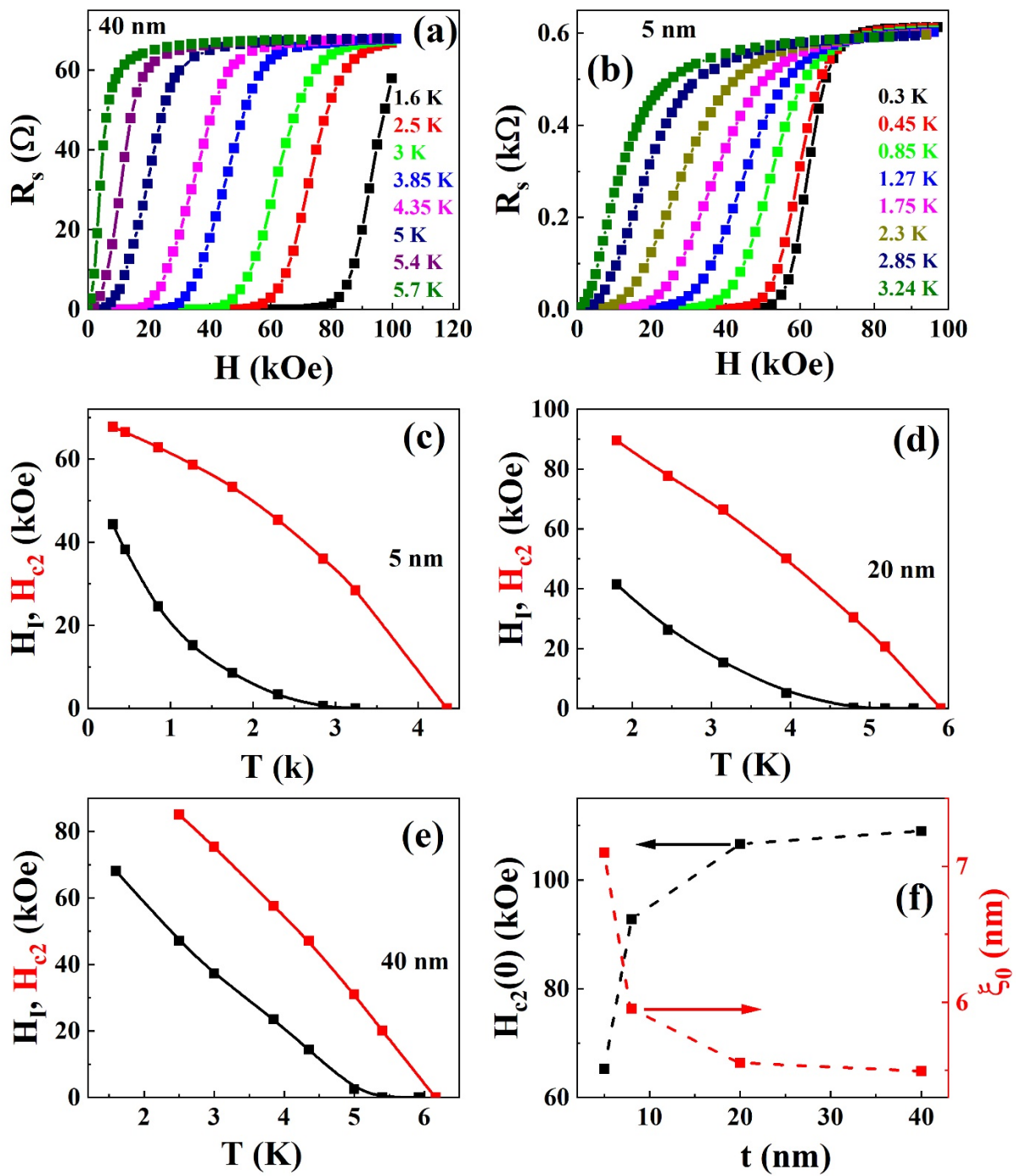

**Figure 4**

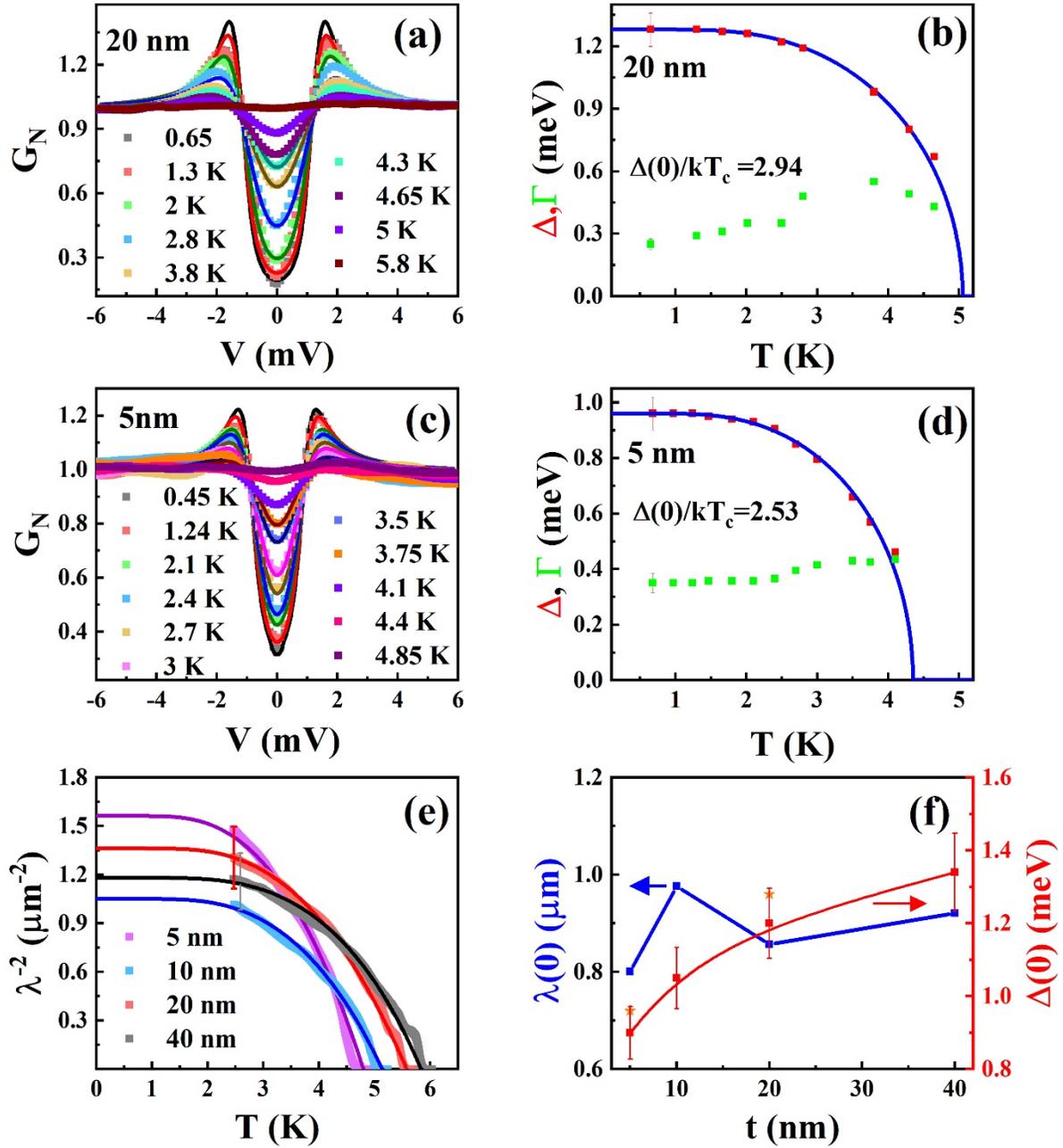

**Figure 5**

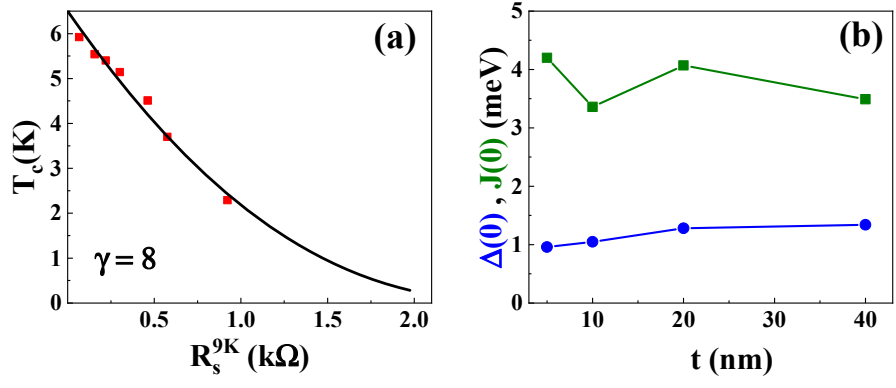